\documentclass[12pt]{article}
\usepackage{mathtools}
\usepackage{graphicx}
\usepackage{amsmath}

\begin{document} 

\title{\bf Generation of Quantum Cluster States using Surface Acoustic Waves} 
\author{MG Majumdar, CHW Barnes} 
\date{\today} 
\maketitle 

\textbf{Abstract.} Cluster State Generation and Use
\\
\\
\\

\section{Introduction}

{ \fontfamily{times}\selectfont
 \noindent One-way quantum computation, also known as Cluster State Quantum Computation, provides a robust and efficient tool to perform universal quantum computation using only single-qubit projective measurements, given a highly entangled cluster state. The cluster-state approach to quantum computation also leads to certain practical advantages such as robustness against errors. 
\\
\\
The cluster state is generated on a basis defined by electrons in Quantum One-Dimensional Channels (Q1DCs), driven by Surface Acoustic Waves [1]. The setup for the generation consists of Copper interdigitated transducers on a Silicon substrate with layers of Silicon Dioxide and Zinc Oxide, to reinforce the piezoelectric effect on the substrate.  The transducers are placed on either sides of a centrally-placed etched region with an Electron Gas. When a high frequency AC signal is applied, Surface Acoustic Waves are generated, by the principle of piezoelectricity. As the SAW propagates through the etched region, the travelling potential it creates carries the electrons from the electron gas with it. 
\\
\\
A typical SAW frequency of 3 GHz and an applied power of 10 dBm produces a measurable current in the nano-ampere range, as shown by \textit{Barnes, et al} [1]. One-qubit rotations and controlled two-qubit gates can be implemented on this system. The primary gate in our generation-protocol is the Root of Swap gate. 
\\
\\
Interchannel and intra-channel, two-instance swap operations form the primary building blocks of the given generation-protocol. \textit{Owen et al} [2] demonstrated how two particles that are interacting in a harmonic potential generate maximally entangled states, which are created simply through the quantum dynamics of the system and possessing a high entanglement ﬁdelity (F $>$ 0.98). The underlying operation is essentially a “root-of-SWAP” operation. \textit{Bayer et al} [3] demonstrated coupling and entanglement of quantum states in a pair of vertically aligned quantum dots by studying the emission of an interacting exciton in a single dot molecule as a function of the separation between the dots. The electron-hole complex was shown to be equivalent to entangled states of two interacting spins. 
\\
\\

\subsection{Surface Acoustic Waves}
{
Surface-acoustic waves (SAWs) are sound waves that travel parallel to the surface of an elastic material. The displacement amplitude decays into the material and therefore these waves are confined to within roughly a wavelength of the surface. In a piezoelectric material, mechanical deformations associated with the SAW produce electric fields.
\\
\\
For non-piezoelectric materials, Hooke's law states that the mechanical stress field experienced by a body is proportional to the strain field:
\\
\\
\begin{center}
$\sigma_{ij}$ = $c_{ijkl}$$\epsilon_{kl}$
\end{center}
where $\sigma_{ij}$ and $\epsilon_{kl}$ are components of the stress and strain fields respectively, and $c_{ijkl}$ is a component of the 4th rank 'elastic' tensor. The electric displacement for non-piezoelectric materials is proportional to components of the electric field, with components of the permittivity tensor being the proportionality constants. 
\\
\\
For piezeolectric materials, the electric displacement depends on the applied electric field and mechanical strain, and the stresses depend on both the applied mechanical strain and the electric field. 
\begin{center}
$D_{i}$ = $\epsilon_{ij}^{S}$ + $e_{ijk}$$\epsilon_{jk}$
\\
$\sigma_{ij}$ = -$e_{kij}$$E_{k}$ + $c_{ijkl}^{E}$$\epsilon_{kl}$
\end{center}
Here superscripts S and E denote that the quantities are measured under constant strain and electric field respectively.
\\
\\
A SAW can be generated by applying an oscillating signal to a suitably designed set of interdigitated transducer based surface gates on a piezoelectric substrate. Small localized displacements of the ﬂuid will propagate as an acoustic wave, also known as a compressional wave. When a SAW passes beneath a SAW transducer of the appropriate pitch, an alternating potential is generated across the transducer.
}
\subsection{The Setup}
{
In our setup, by bringing the channels close to each other, we allow for Coulombic interaction to take place between the electrons travelling in the channels. As seen, with a high fidelity, this generates an entangled state using the 'Root-of-Swap' operation. One can also use a magnetic field, oriented in a certain direction to implement single qubit rotations, which constitute an essential part of the Universal Quantum Gates set. 
\begin{center}
\scalebox{0.55}{\includegraphics{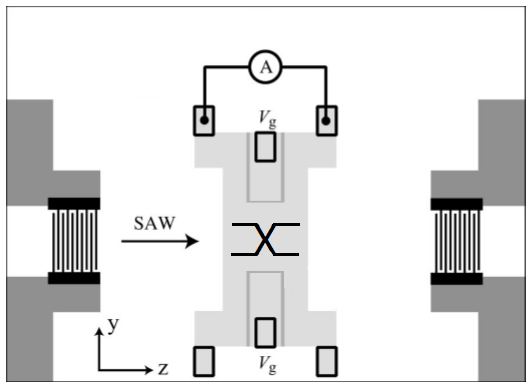}}
\\
\textbf{Figure 1}: \textit{Setup, comprising of Quantum One-Dimensional Channels with electrons driven by Surface Acoustic Waves (SAWs)}
\end{center}
}

\section{Characterization of Entanglement}

We wish to characterize the entanglement in multipartite qubit states. A pure n-qubit state is called unentangled if its wave function may be written as an n-fold tensor product of individual qubits. A state is globally entangled if it cannot be written as a tensor product of any set of subsystems.
\\
\\
There are several ways of quantifying entanglement. Measures of entanglement can be been used that are constant on locally equivalent states.  These must be entanglement monotones i.e. they must be non-increasing under Local Operations and Classical Communication (LOCC). 
\\
\\
One can also have observables whose expectation values are positive (negative) on unentangled states and negative (positive) on entangled states. 

\subsection{Partial Density Matrices}
{
The density operator $\rho$ for the ensemble or mixture of states  $\vert\psi_{i}\rangle$        with probabilities   $p_{i}  $  is given by
\\
\begin{center}
$\rho$ = $\sum_{i}$$p_{i}$$\vert\psi_{i}\rangle$$\langle\psi_{i}\vert$
\end{center}
The reduced density operator describes the properties of measurements of a sub-system A, when the other subsystem(s) is(are) left unobserved, by tracing them out.
\\
\\
Peres [4] showed that a necessary condition for seperability in a system is that a matrix obtained using partial transposition of the density matrix of the system has only non-negative eigenvalues.
}
\subsection{Concurrence}
{
As defined by \textit{Carvalho et al} [5], for an N-partite quantum system, one can define $2^{N}-2$ reduced density matrices and an associated concurrence measure: 
\begin{center}
$C_{N}$ = $2^{1-\frac{N}{2}}$$\sqrt{(2^{N}-2){(\langle\psi\vert\psi\rangle)}^{2} - \sum\limits_{\alpha}{Tr\rho_{\alpha}}^{2}}$
\end{center}
where $\alpha$ labels all the reduced density matrices. 
\\
}

}

\section{Results and Discussion}
\label{sec:MLE}

{ \fontfamily{times}\selectfont
 \noindent
\begin{center}
\scalebox{0.6}{\includegraphics{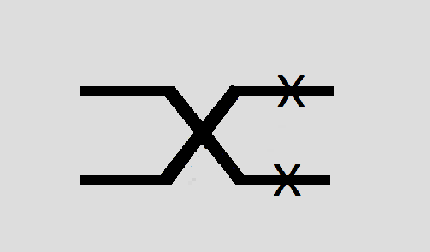}}
\\
\textbf{Figure 2}: \textit{Gate Combination with interchannel and intrachannel Root-of-Swap Operations}
\end{center}
For this setup, we consider the various input states and the concurrence measures for the entanglement generated by the setup in the process. 
\\
\\
\textbf{Case 1}: Input comprises of $\vert00\rangle$ and $\vert00\rangle$
\begin{center}
$C_{4}$ = 0
\end{center}
\textbf{Case 2}: Input comprises of $\vert11\rangle$ and $\vert11\rangle$
\begin{center}
$C_{4}$ = 0
\end{center}
Both these cases are expected to have vanishing entanglement concurrence-measures, as the Root-of-Swap operation leaves the $\vert11\rangle$/$\vert00\rangle$ combinations unaltered. In this case, a seperable input composite state is unaffected by the \textit{setup-entanglers}.
\\
\\
\textbf{Case 3}: 
\\
\indent Input comprises of $\vert00\rangle$ and $\vert01\rangle$ \\
\indent Input comprises of $\vert00\rangle$ and $\vert10\rangle$ \\
\indent Input comprises of $\vert01\rangle$ and $\vert00\rangle$ \\
\indent Input comprises of $\vert10\rangle$ and $\vert00\rangle$ \\
\indent Input comprises of $\vert11\rangle$ and $\vert10\rangle$ \\
\indent Input comprises of $\vert11\rangle$ and $\vert01\rangle$ \\
\indent Input comprises of $\vert10\rangle$ and $\vert11\rangle$\\
\indent Input comprises of $\vert01\rangle$ and $\vert11\rangle$\\
\begin{center}
$C_{4}$ = 1.479
\end{center}
In these cases, there is one flipped spin, with respect to the remaining qubit subsystem. As a result, the entanglement capacity for each of these systems is equal. 
\\
\\
\textbf{Case 4}:
\\
\indent Input comprises of $\vert00\rangle$ and $\vert11\rangle$\\
\indent Input comprises of $\vert11\rangle$ and $\vert00\rangle$
\\
\\

\begin{center}
$C_{4}$ = 1.458
\end{center}
The first inter-channel entangling Root-of-Swap operation has no effect on the input state since they are $\vert11\rangle$/$\vert00\rangle$ combinations. However, the subsequent intra-channel entanglers give rise to entanglement in the state. 
\\
\\
\textbf{Case 5}: \\
\indent Input comprises of $\vert01\rangle$ and $\vert01\rangle$\\
\indent Input comprises of $\vert10\rangle$ and $\vert10\rangle$
\\
\\

\begin{center}
$C_{4}$ = 1.620
\end{center}
This is the case when both interchannel and intrachannel entanglers contribute to the generation of entanglement. 
\\
\\
\textbf{Case 6}:  \\
\indent Input comprises of $\vert01\rangle$ and $\vert10\rangle$ \\
\indent Input comprises of $\vert10\rangle$ and $\vert01\rangle$
\\
\\
\begin{center}
$C_{4}$ = 1.225
\end{center}
This is an interesting case wherein the entanglers contribute to the generation of entanglement, much like in \textit{Case 5}. However, the concurrence measure is much lower in this case. 
\begin{center}
\scalebox{0.25}{\includegraphics{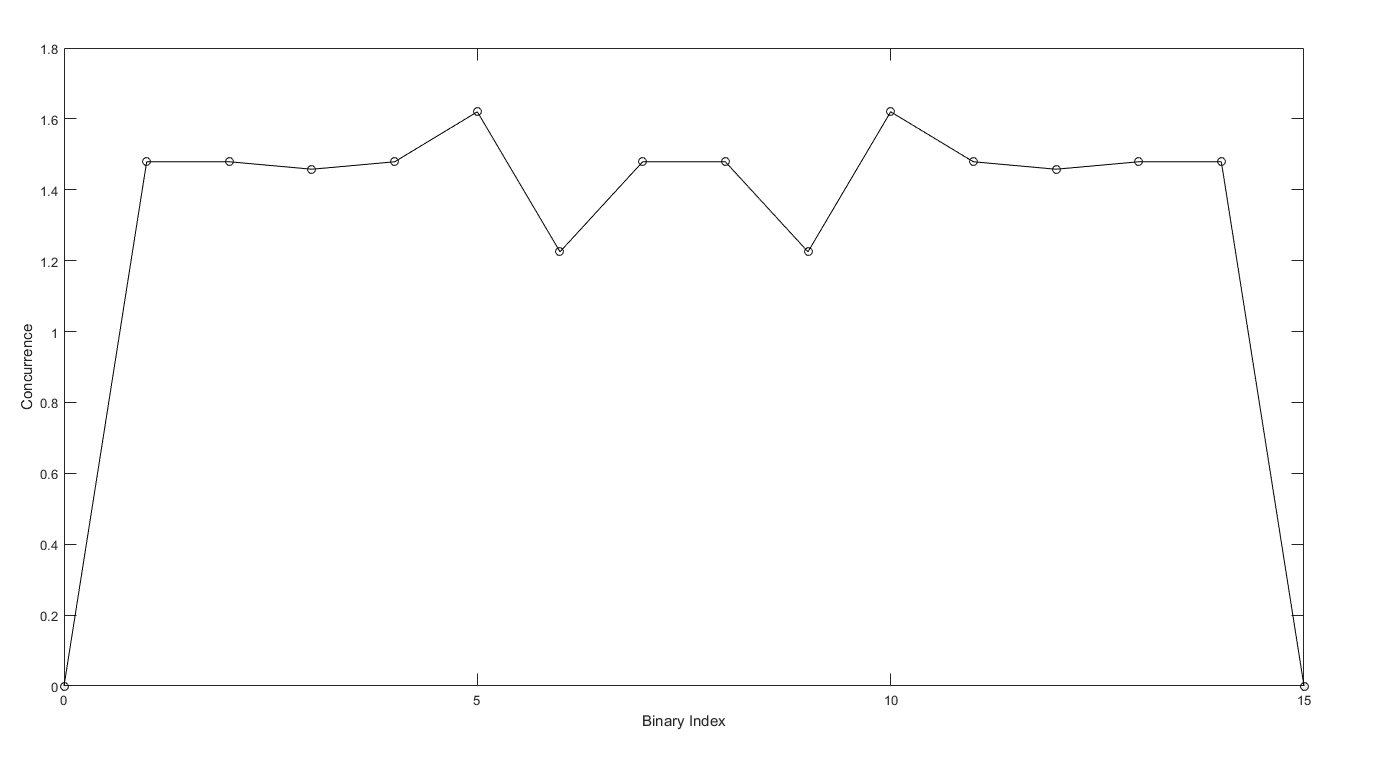}}
\\
\textbf{Figure 3}: \textit{Concurrence Plot}
\end{center}
We hypothize that the dip in the plot (\textit{Case 6}) is because of the concept of \textit{Entanglement Monogamy}. Once the entanglement is generated by the interchannel entanglers, the intrachannel entanglers entangle the states further, though this essentially reduces entanglement between subsystems and we obtain a cluster state. 
\begin{center}
\scalebox{0.5}{\includegraphics{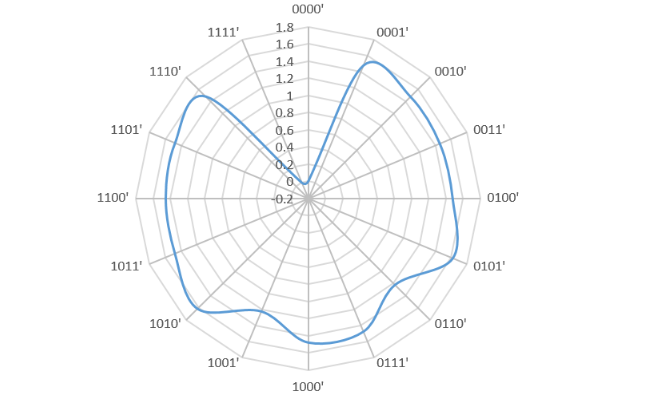}}
\\
\textbf{Figure 4}: \textit{Polar Plot of Concurrence Measures}
\end{center}

\begin{center}
\scalebox{0.6}{\includegraphics{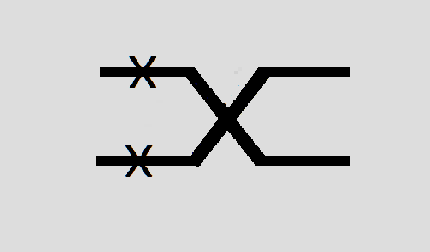}}
\\
\textbf{Figure 5}: \textit{Gate Combination with intrachannel and interchannel Root-of-Swap Operations}
\end{center}

In this case, we have vanishing concurrence (implying seperability) for input comprising of $\vert00\rangle$$\vert00\rangle$ and $\vert11\rangle$$\vert11\rangle$, as in the previous setup. 
\\
\\
This is due to the entanglers not generating entanglement for this particular input state, given a Root-of-Swap based generator setup, irrespective of the order of the entanglers: first intra- and then inter-channel entanglers, or first inter- and then intra-channel entanglers.  
\\
\\
For the case with one spin flipped, with respect to other qubits in the system, we have the same result as for the previous setup. The concurrence remains at 1.479. The concurrence measure and the amount of entanglement remains unchanged due to the fact that after the first entangler operation in both circuits, entanglement is generated only in a single two-qubit subsystem while the remaining two-qubit subsystem remains in a composite state. This step remains unchanged due to the symmetry of this particular kind of four-qubit input state. Eventually, the second entangler generates entanglement in the entire system by generating quantum correlations between one part of the entangled two-qubit subsytem and one half of the composite subsystem of qubits. 
\begin{center}
\scalebox{0.5}{\includegraphics{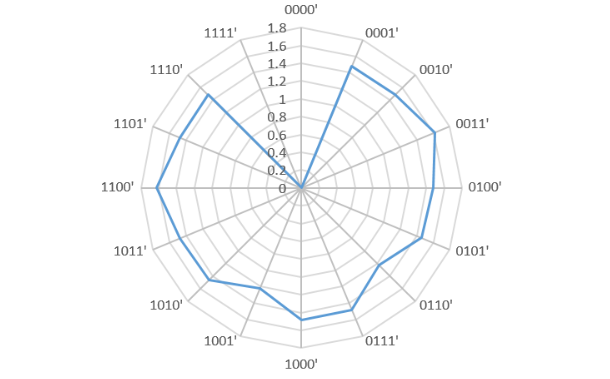}}
\\
\textbf{Figure 6}: \textit{Polar Plot of Concurrence Measures}
\end{center}
The concurrence for the $\vert00\rangle$$\vert11\rangle$/$\vert11\rangle$$\vert00\rangle$ and the $\vert01\rangle$$\vert01\rangle$/$\vert10\rangle$$\vert10\rangle$ states are interchanged, with respect to the case for the inter-intrachannel combination. The former has a concurrence of 1.620 while the latter has a concurrence of 1.458. This is because the switch in the entangler combination and sequence is countered by the rearrangement of input qubits for the respective matching concurrence measures in the two setup-cases. For the $\vert01\rangle$$\vert10\rangle$/$\vert10\rangle$$\vert01\rangle$, the concurrence remains at 1.225.

\begin{center}
\scalebox{0.6}{\includegraphics{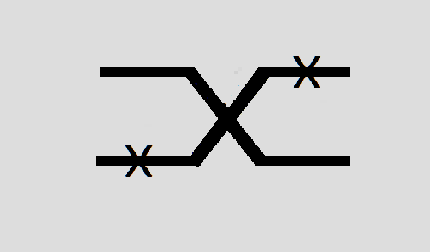}}
\\
\textbf{Figure 7}: \textit{Special Gate Combination ('Cross-Arm Mobius')}
\end{center}
In this case, for same-spin qubit input, concurrence vanishes, while for the case with one spin flipped, with respect to other qubits in the system, we have a higher concurrence than the previous case. The value of concurrence for this input combination and the setup (Figure 7) is 1.571. 
\\
\\
The concurrence for the $\vert00\rangle$$\vert11\rangle$/$\vert11\rangle$$\vert00\rangle$ and the $\vert01\rangle$$\vert01\rangle$/$\vert10\rangle$$\vert10\rangle$ states are higher or equal to the previous setup-cases. The former has a concurrence of 1.894 while the latter has a concurrence of 1.620.
\\
\\
 For the $\vert01\rangle$$\vert10\rangle$/$\vert10\rangle$$\vert01\rangle$, the concurrence value is 1.785. The possible cause for higher concurrence for all input combinations is viewed in the entanglement within the various subsystem partitions. Previously, there was a trade-off between the contribution of an entangled partition-class and the seperability of remaining subsystem partition-classes. In this setup, the entanglement is present across the various partitions and subsystems. Thus this setup, named as the 'Cross-Arm Mobius', is a good generator of entanglement in SAW-driven electrons. 
\\
\begin{center}
\scalebox{0.4}{\includegraphics{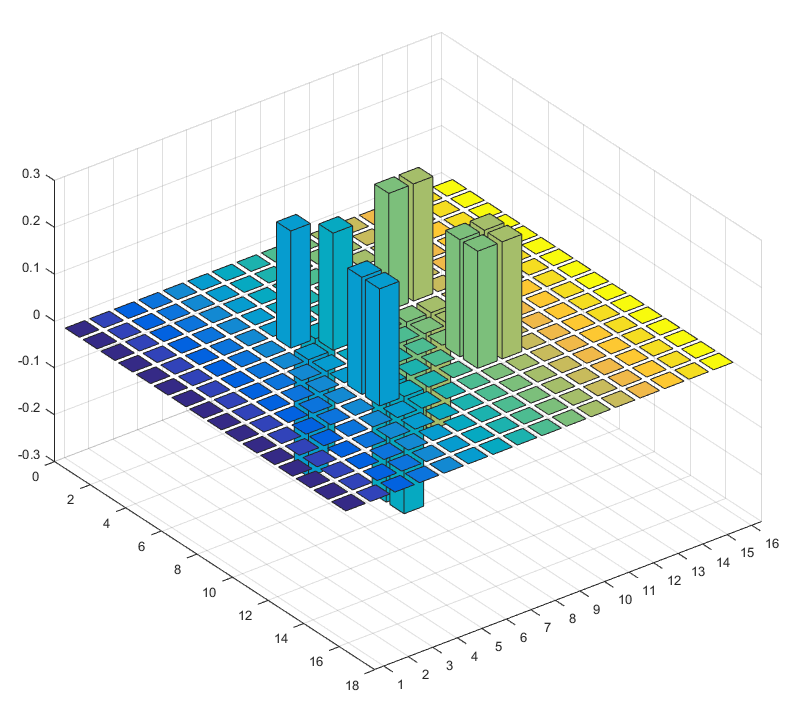}}
\\
\textbf{Figure 8}: \textit{Density Matrix for $\vert0110\rangle$ case and inter-intrachannel setup}
\end{center}

\begin{center}
\scalebox{0.4}{\includegraphics{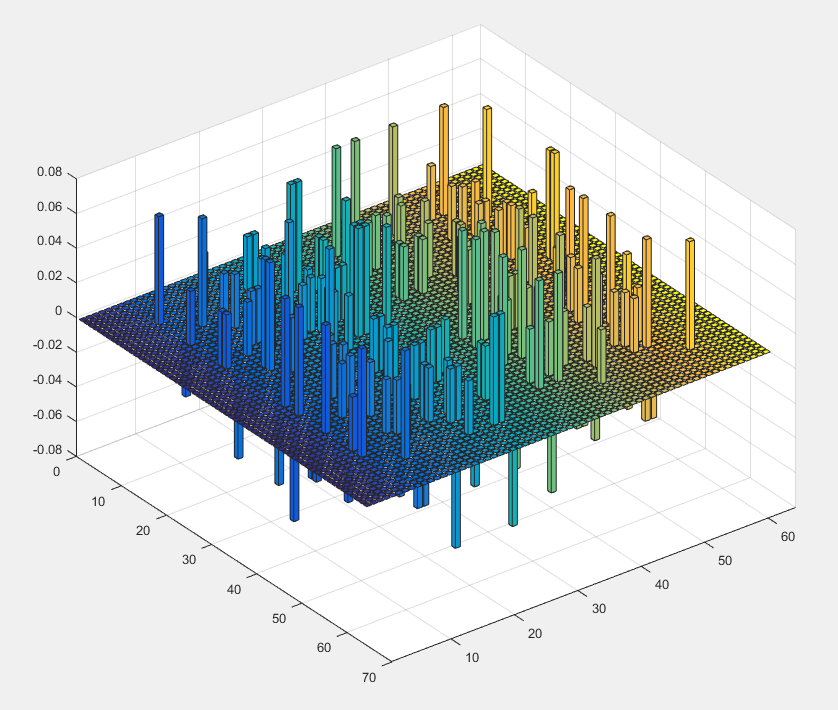}}
\\
\textbf{Figure 9}: \textit{Density Matrix for $\vert010101\rangle$ case with given protocol for generation of Cluster States}
\end{center}

\section{Conclusion}

We have developed a scheme for the generation of entanglement and cluster states on a basis defined by electrons in Quantum One Dimensional Channels (Q1DCs), driven by Surface Acoustic Waves (SAWs). We have characterized the states obtained to quantify entanglement in the states and to devise optimal circuits for generation of entangled multipartite qubit-states.
\\

\section{Acknowledgements}

We would like to thank the Nehru Trust for Cambridge University (NTCU), Trinity College - Cambridge, the Hitachi Cambridge Laboratory and the Thin Film Magnetism group of Cavendish Laboratory, University of Cambridge for their support in the pursuit of this project.

[1] Barnes, C. H. W., J. M. Shilton, and A. M. Robinson. "Quantum computation using electrons trapped by surface acoustic waves." Physical Review B 62.12 (2000): 8410.
\\
\\
{[2]} Owen, E. T., M. C. Dean, and C. H. W. Barnes. "Generation of entanglement between qubits in a one-dimensional harmonic oscillator." Physical Review A 85.2 (2012): 022319.
\\
\\
{[3]} Bayer, M., et al. "Coupling and entangling of quantum states in quantum dot molecules." Science 291.5503 (2001): 451-453.
\\
\\
{[4]} Peres, Asher. "Separability criterion for density matrices." Physical Review Letters 77.8 (1996): 1413.
\\
\\
{[5]} Carvalho, André RR, Florian Mintert, and Andreas Buchleitner. "Decoherence and multipartite entanglement." Physical review letters 93.23 (2004): 230501.
\\
\\
{[6]} Rungta, Pranaw, et al. "Universal state inversion and concurrence in arbitrary dimensions." Physical Review A 64.4 (2001): 042315.
\\
\\
{[7]} Hill, Scott, and William K. Wootters. "Entanglement of a pair of quantum bits." Physical review letters 78.26 (1997): 5022.
\\
\\
{[8]} Bennett, Charles H., et al. "Mixed-state entanglement and quantum error correction." Physical Review A 54.5 (1996): 3824.
\\
\\
{[9]} Horodecki, Michał, Paweł Horodecki, and Ryszard Horodecki. "Limits for entanglement measures." Physical Review Letters 84.9 (2000): 2014.
\\
\\
{[10]} Wong, Alexander, and Nelson Christensen. "Potential multiparticle entanglement measure." Physical Review A 63.4 (2001): 044301.
\\
\\

\end{document}